\documentclass{jfm}
\usepackage{graphicx}
\usepackage{bm}
\usepackage{amsmath}
\usepackage{float}
\usepackage{amssymb}
\usepackage[amssymb]{SIunits}
\usepackage[dvipsnames]{xcolor}
\usepackage{tabularx}
\usepackage{booktabs}
\usepackage{soul}
\usepackage{cite}
\usepackage{multirow}
\usepackage[colorlinks,citecolor = black, linkcolor= blue,hyperindex,CJKbookmarks]{hyperref}

\title{Evaporation of acoustically levitated bicomponent droplets: mass and heat transfer characteristics}

\author{Yuki Wakata \aff{1}, 
Xing Chao \aff{1}, 
Chao Sun \aff{1,2}
\corresp{\email{chaosun@tsinghua.edu.cn}}
Christian Diddens \aff{3}}

\affiliation{
	\aff{1}New Cornerstone Science Laboratory, Center for Combustion Energy, Key Laboratory for Thermal Science and Power Engineering of Ministry of Education, Department of Energy and Power Engineering, Tsinghua University, 100084 Beijing, China

	\aff{2}Department of Engineering Mechanics,
	School of Aerospace Engineering, Tsinghua University, 100084 Beijing, China

    \aff{3}Physics of Fluids, University of Twente, 7522 NB Enschede, The Netherlands	
}

\begin{document}
\maketitle

\begin{abstract}
Evaporation of multicomponent droplets is important in a wide range of applications, albeit complex, and requires a careful investigation.
We experimentally and numerically investigate the evaporation characteristics of spherical, ethanol-water droplets with different initial concentration ratios in the acoustic levitation field.
Imaging techniques and infrared thermometry are used for acquiring volume and surface temperature variations of droplets, reflecting their mass and heat transfer characteristics. 
Numerical simulations are conducted using modified parameters based on a theoretical model to consider the effect of the acoustic field. 
The calculation results show good agreement with the experimental data.
The concentration and temperature distribution within the droplet is further investigated based on the numerical results.

\end{abstract}


\section{Introduction}
Evaporation of spray droplets is ubiquitous in many natural and industrial processes, including spray cooling, inkjet printing, fuel combustion and even virus transmission via saliva droplets \citep{Zang2019, Lieber2021, Lohse2022fundamental}.
From the most elementary model proposed by \citet{Maxwell1877}, which considers the purely diffusive evaporation of a single spherical droplet, researches refined various aspects of this problem to be more relevant to reality, considering the effects of the Stefan flow \citep{Fuchs1959}, non-spherical droplets \citep{Tonini2013exact, Al2018}, gas temperature gradient \citep{Tonini2012analytical}, etc. 
When the system involves two or more components, the evaporation process becomes significantly more complex due to the introduction of additional factors, including the coupling of different component evaporation, thermal and solutal Marangoni flow generated by local concentration gradients, vapour condensation, etc \citep{Law1982, Sirignano1983, Li2023high, Tonini2019analytical}. 
The evaporation of multicomponent droplets placed on a surface has been extensively investigated by studies such as \citet{Diddens2017evaporating, Diddens2021competing}, and is summarised in a recent review by \citet{Wang2022}. 
Nevertheless, experimental studies of non-contact spray droplet evaporation are limited due to experimental methodology constraints.
Different methods have been designed to investigate droplet evaporation in mid-air, such as directly measuring sprays generated by nozzles \citep{Li2023high}, or using monodisperse droplet streams generated through Rayleigh-type disintegration \citep{Maqua2008Monodisperse}.

The acoustic levitation method has received increasing attention due to their ability to stably levitate the sample fluid of a considerable size for an extended duration without any surface contact \citep{Zang2017Acoustic, Andrade2018Review,OConnell2023}. 
The steady-state geometry and spatial location of an acoustically levitated droplet have been well studied both theoretically and experimentally \citep{Tian1993, Shi1996, Yarin1998}, while further research is required to understand its evaporation characteristics, which are influenced by the secondary flows introduced by the acoustic field, known as acoustic streaming \citep{Riley2001, Lee1990}. 
Such acoustic streaming includes strong convection close to the droplet interface (inner acoustic streaming) and large-scale toroidal vortices (outer acoustic streaming), which were observed experimentally by \citet{Trinh1994}. 

\citet{Yarin1999} proposed a theoretical solution for the flow field around an acoustically levitated droplet of pure liquid and the corresponding Sherwood number (the ratio of convective to diffusive mass transfer) for the droplet evaporation.
Their theoretical solution has been extended to the evaporation of bicomponent droplets in \citet{Yarin2002} and have been utilized in a wide range of studies \citep{Al2011, Chen2022, Zeng2023}. \citet{Brenn2007}, on the other hand, applied \citet{Lierke1996}'s model in conjunction with correlation functions to calculate the Sherwood number and examine the evaporation of multicomponent droplets. Numerical simulations for acoustically levitated droplets have also been investigated \citep{Bansch2018numerical, Doss2022}, but there is currently a lack of studies on multicomponent droplets, and validation with experimental results is also lacking.
As the mass and heat transfer are coupled in the evaporation process \citep{Sirignano2010fluid}, additional attention should to be paid to the droplet temperature variation and to perform a quantitative analysis of heat transfer characteristics in the acoustic field.
In this regard, \citet{Sasaki2020} measured the temperature variation of bicomponent droplets and obtained the heat transfer coefficients. Further investigation is needed to interpret these coefficients mechanistically and investigate the coupling of heat and mass transfer.
Moreover, it is worth noting that some studies (e.g. \citet{Yarin1999, Yarin2002, Al2011}) utilised ventilation flow to eliminate vapour accumulation in the vortices of outer acoustic streaming, while some recent studies \citep{Chen2022, Doss2022, OConnell2023, Zeng2023} have not introduced this external flow. The effect of the absence of the ventilation flow on evaporation should be sorted out and there needs a model that allows for calculations in both situations with and without a ventilation flow.

In our work, we experimentally and numerically investigate the coupling effect of the mass and heat transfer in the evaporation of acoustically levitated water-ethanol droplets.
We comprehensively measured the variation of volume and surface temperature of droplets evaporating at low humidity (5\%) and without ventilation flow blowing off the outer streaming vortices.
A numerical model using finite element methods is developed to simulate the evaporation process of a levitated spherical droplet.
We derived an extended model to include the effects of the outer acoustic streaming and added them to the numerical simulations through modified Sherwood and Nusselt numbers.
The calculated results are compared with the experimental results to verify the accuracy. At the same time, the calculated results provide information on the concentration and temperature distribution of the droplets.

The paper is organized as follows: 
we first give details of the experimental method and materials in Section §\ref{sec:exp}. 
In Section §\ref{sec:cal}, we present our computational model for the evaporation of acoustic levitated droplets. Results related to the mass and heat transfer characteristics will be described in Section  §\ref{sec:results}, including comparisons between the experiment and the model. 
The paper ends with conclusions and an outlook.

\begin{figure}
\centering
\includegraphics[width=0.95\linewidth]{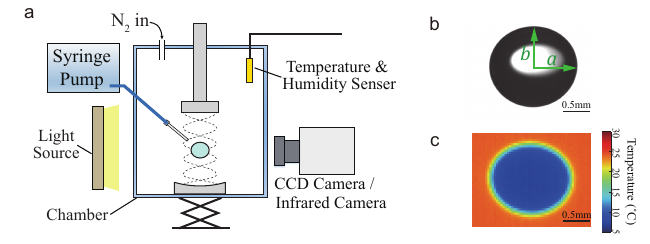}
  \caption{(\textit{a}) Schematic of the experimental setup for studying the evaporation of an acoustically levitated droplet at controlled temperature and humidity. (\textit{b}) Image of the levitated droplet obtained by the CCD camera, where $a$ is the longer axis and $b$ is the shorter axis of the projected ellipse.  (\textit{c}) The temperature field obtained by the infrared camera. The scale bar relates to $\unit{0.5}{\milli\meter}$.}
\label{fig:setup}
\end{figure}

\section{Experiments and materials}
\label{sec:exp}
The experimental setup of the current study is illustrated in figure \ref{fig:setup}(a).
A coaxial acoustic levitator, consisting of a piezocrystal transducer (operating frequency 20.5 kHz) and a reflector, generates a standing acoustic wave with five nodes.
In the experiment, the sample fluid is placed at the middle pressure node and is forced upwards by the acoustic radiation pressure to counterbalance the gravity force.
The droplet is generated by a syringe pump (Harvard Apparatus PHD ULTRA) with a 15-gauge stainless steel needle. The initial volume of the droplet is controlled by the flow rate and the injection time to be $\unit{1.481}{\milli\meter^3}$, which is equivalent to an effective diameter square of $D^2=\unit{2}{\milli\meter^2}$. 
The temperature of the droplets released from the needle will be lower than the ambient temperature due to the evaporative cooling during the generation process. This temperature reduction is justified to have a negligible effect on the experimental results and the variation in droplet composition during the droplet generation process is also negligible.
The levitated droplet has an oblate spheroidal shape, with an aspect ratio of the droplet contour $k=a/b$, where $a$ is the longer axis and $b$ is the shorter axis of the projected ellipse (see figure \ref{fig:setup}(b)).
For the present study, the aspect ratio is limited to less than 1.2 to take advantage of the small sphere assumption \citep{Yarin1998, Yarin1999}. 

The experiments are conducted in a closed chamber with nitrogen flow input to maintain the ambient humidity at 5\%.
The humidity is controlled at a very low level during the evaporation process to minimise the effects of water vapour condensation \citep{Li2023high}.
Once the humidity reaches 5\%, the nitrogen input flow rate is reduced to maintain a basic positive pressure, preventing the entry of high humidity air. The gas temperature in the chamber is controlled at $\unit{25\pm0.5}{\degree C}$. During the experiments, a sensor (KS-SHTE1KT, Keshun Ltd.) with an accuracy of $\unit{\pm 0.2}{\degree C}$ for temperature and $\unit{\pm 2 \%}{RH}$ for humidity is used to sample the temperature and the humidity of the gas. 

A CCD camera (Ximea XiD) captures the volume variation of the droplet during evaporation using backlighting, and the surface temperature variation of the droplet is measured by an infrared camera (Telops FAST L200) through a high-transmittance germanium glass window (transmittance=0.95).  Figure \ref{fig:setup}(\textit{b}) and Figure \ref{fig:setup}(\textit{c}) display the captured CCD and infrared camera images, respectively. 
The infrared measurement is calibrated using liquid surfaces with temperatures controlled by a water bath.
The objective temperature when measuring droplet's surface temperature is determined as the minimum temperature within the area of the droplet. This methodology is in line with previous articles \citep{Sasaki2020}. 

Droplets used in the experiments are binary mixtures composed of ultrapure water (prepared by a Milli-Q purification system from Merck, Germany) and ethanol (General-Reagent, 99.7\%). Solutions of different volume fractions of ethanol (0\%, 25\%, 50\%, 75\%, 100\%) are prepared at a constant room temperature of $\unit{25}{\degree C}$.
Table \ref{Tab1} presents the physicochemical properties of the two components at $\unit{10}{\degree C}$ and $\unit{25}{\degree C}$.

 \begin{table}
	\centering
	\renewcommand\arraystretch{1.5}
	 	\setlength{\tabcolsep}{4mm}{
		\begin{tabular}{cccccc}
               &                             & \multicolumn{2}{c}{Water} & \multicolumn{2}{c}{Ethanol} \\  
	               &                             & $\unit{10}{\degree C}$         & $\unit{25}{\degree C}$         & $\unit{10}{\degree C}$         & $\unit{25}{\degree C}$           \\
	Liquid density $\rho$  & $\rm{kg /m^3}$     & 999.7 & 997.0 & 798.1 & 785.0 \\
	Latent heat $h_{fg}$            & $\rm{kJ/kg}$                       & 2477.2      & 2441 .7      & 1043.8      & 1025.5      \\
	Saturation pressure $P_{sat}$           & $\rm{kPa}$                         & 1.23      & 3.17      & 3.14      & 7.89        \\
	Diffusion coefficient $D$ & $\rm{10^{-5}\ m^2/s }$ & \multicolumn{2}{c}{2.5}                       & \multicolumn{2}{c}{1.35}                    \\ 
	Molecular weight $MW$ & g/mol  & \multicolumn{2}{c}{18.02}                       & \multicolumn{2}{c}{46.07}                    \\ 
		\end{tabular}}
  \caption{Properties of test liquids at $\unit{10}{\degree C}$ and $\unit{25}{\degree C}$.}
\label{Tab1}
\end{table} 
 
\section{Calculation model}
\label{sec:cal}
In this section, we present our calculation model for simulating the evaporation of an acoustically levitated bicomponent droplet. 
Since the aspect ratio $k$ of the droplet shape is close to 1, we assume a spherically symmetric droplet with radius $R_d$ and neglect the internal circulation flow in the droplet.
This assumption is not valid for droplets with aspect ratio $k>1.25$, when droplets are ellipsoidal and strong internal flows are experimentally observed \citep{Yarin1999, Yamamoto2008, Sasaki2019}.
The effect of the acoustic field on droplet evaporation is considered by introducing modified Nusselt and Sherwood numbers for heat and mass transfer, which will be presented in Section \ref{Acoustic}.

\subsection{Heat and mass transfer equations}
The calculation is performed using the an in-house finite element framework based on \textsc{oomph-lib} \citep{Heil2006} for droplet evaporation, which takes into account both species and temperature gradients within the droplet.
A comprehensive explanation of this finite element model can be found in \citet{Diddens2017evaporating} and \citet{Diddens2017detailed}.

In the gaseous phase, the distribution of the vapour mass fraction $c_i$, with $i=w$ for water and $i=e$ for ethanol, is described using the following modified diffusion equation that considers the effect of convection in the acoustic field by introducing the effective Sherwood number ${S\!\!\: h}_{e\!f\!\!f, i}$:
\begin{equation}
\frac{\partial c_i}{\partial t}=\frac{{S\!\!\: h}_{e\!f\!\!f, i}}{2}D_{i}^g\ \frac{1}{r^2} \frac{\partial}{\partial r}\left(r^2 \frac{\partial c_i}{\partial r}\right),
\label{MassDiff}
\end{equation}
Here, $D_{i}^g$ is the mass diffusion coefficient of component $i$ in the gaseous phase. The ${S\!\!\: h}_{e\!f\!\!f, i}$ here represents an average effect of the acoustic streaming convection on the concentration field.When ${S\!\!\: h}_{e\!f\!\!f, i}=2$, equation (\ref{MassDiff}) is equivalent to the equation for pure diffusive evaporation. The calculation of  ${S\!\!\: h}_{e\!f\!\!f, i}$ will be detailedly illustrated in Section \ref{Acoustic}.
To solve the equation (\ref{MassDiff}), boundary conditions at the liquid-gas interface $r=R_d$ and at $r\rightarrow \infty$ have to be imposed. 
The vapour concentration at the interface, $c_{s,i}$, is calculated as:
\begin{equation}
c_{s,i}=\gamma_i x_{i}^l \frac{MW_i\ p_{sat, i}}{R T_s},
\label{interface}
\end{equation}
Here, $x_{i}^l$ is the mole fraction of the liquid phase component $i$, $MW_i$ the molecular weight, $p_{sat, i}$ the saturation vapour pressure of the pure component $i$ as a function of $T_s$, which is the temperature evaluated at the interface. $\gamma_i$ is the activity coefficient calculated by the thermodynamic model AIOMFAC \citep{Zuend2008}.
At far field ($r\rightarrow \infty$), the vapour concentration of water relates to the humidity $\phi$ of the surrounding gas:
\begin{equation}
c_{\infty, w}=\phi \frac{MW_w p_{sat, w}}{R T_{\infty}},
\label{cinfty}
\end{equation}
while there is no ethanol in the surrounding gas, i.e. $c_{\infty, e}=0$.

In the liquid phase, the mass fraction of the components $y_i^l$ is governed by the following diffusion-convection equation:
\begin{equation}
\rho^l \left( \frac{\partial y_i^l}{\partial t} + u \cdot \frac{\partial y_i^l}{\partial r}\right)=\frac{1}{r^2} \frac{\partial}{\partial r}\left(r^2 \rho^l D_{i}^l \frac{\partial y_i^l}{\partial r}\right)-J_i^l \cdot  \delta_{l g}
\end{equation}
Here, $\rho_l$ is the liquid density, which is allowed to dependent of the composition. Due to the changing mass density of the liquid mixture during the evaporation process, the radial velocity $u$ can be non-zero and in order to conserve the species masses, the radial advection term $u \cdot \frac{\partial y_i^l}{\partial r}$ must be considered. The radial velocity  $u$ is calculated to be less than less than $\unit{5\times10^{-8}}{\meter/\second}$ throughout the evaporation process and has little effect on the specific mass and heat transfer.
$D_{i}^l$ is the diffusion coefficient in the liquid phase and the $J_i^l \cdot \delta_{lg}$ term represents a source/sink term at the droplet interface, where $\delta_{l g}$ is the interface delta function and $J_i^l$ the diffusive flux in the liquid phase calculated by (3.7) in \citet{Diddens2017evaporating}.

During evaporation, the droplet gets cooled down due to the evaporation at the interface, which in turn affects the evaporation rate by the virtue of (\ref{interface}). At the same time, the temperature field of the gas may also vary. To study this, we calculate the temperature distribution in both the liquid and gas phase using the following energy equation:
\begin{equation}
\rho c_p\left( \frac{\partial T}{\partial t} + u \cdot \frac{\partial T}{\partial r}\right)=\frac{1}{r^2} \frac{\partial}{\partial r}\left(r^2 \lambda \frac{\partial T}{\partial r}\right)-\left(j_w^{l g} \Lambda_w+j_e^{l g} \Lambda_e\right) \cdot \delta_{l g} ,
\label{temp}
\end{equation}
where the density $\rho$, the specific heat capacity $c_p$ and the thermal conductivity $\lambda$ are differentiated in the gas phase and liquid phase. $j_e$ and$j_w$ are the mass transfer rates of ethanol and water, respectively, and $\Lambda_e$ and $\Lambda_w$ denote the corresponding latent heat of evaporation. The $\lambda$ in gaseous phase is adjusted by a Nusselt number $Nu$ to consider the effect of acoustic streaming, i.e. $\lambda_g=\frac{{N\!\!\: u}_{e\!f\!\!f}}{2}\lambda_{air}$, where $\lambda_{air}$ is the thermal conductivity of air. The calculation of the Nusselt number will be introduced in the next subsection.

The boundary and initial conditions of the calculation problem are set as follows:
\begin{equation}
\text{At infinity } (r=\infty): c_w=c_{\infty,w} \text{ from equation(\ref{cinfty})},\  c_e=0,\  T=T_0=\unit{25}{\degree C}.\\
\end{equation}
 
Regarding the initial conditions, as stated in section \ref{sec:exp}, the generation process of the droplet has a negligible effect on the experimental results, so we assume that the droplet evaporates from room temperature ($\unit{25}{\degree C}$) and the set volume concentration. Defining the set volume concentration of the component $i$ in the liquid mixture as $\varphi_i$. When $t=0$:
\begin{equation}
\text{In liquid phase }(0<r<R_d): \  y_i^l=\frac{\rho^l_i \varphi_i}{\rho^l_1 \varphi_1+\rho^l_2 \varphi_2}, \ T=T_0. \\
\end{equation}
\begin{equation}
\text{In gaseous phase }(R_d<r<\infty): \   c_w=c_{\infty,w} , \  c_e=0, \  T=T_0. \\
\end{equation}

\subsection{Effect of acoustic streaming}
\label{Acoustic}
\begin{figure}
\centering
\includegraphics[width=1\linewidth]{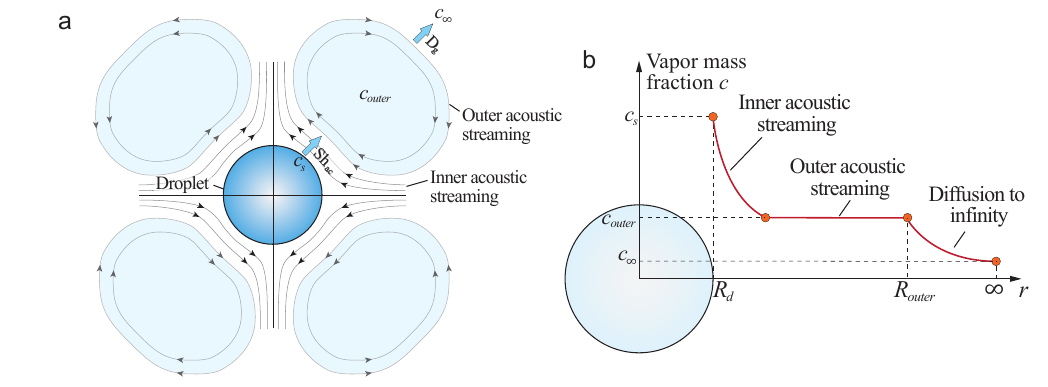}
\caption{\textbf{The evaporation model:} (\textit{a}) Schematic of the flow field outside the acoustic levitated droplet . (\textit{b}) Distribution of the vapour mass fraction $c$ with radius $r$. }

\label{fig:model}
\end{figure}

Acoustic levitation is known to generate secondary flow circulations around the droplet, which is called acoustic streaming. A detailed theoretical analysis of this secondary flow is provided by \citet{Yarin1998, Yarin1999}. Consequently, the acoustic field influences the droplet's evaporation in two ways: The introduction of convection at the acoustic boundary layer close to the droplet (inner acoustic streaming), and the formation of large-scale toroidal vortices about the droplet (outer acoustic streaming), which affects the far-field conditions. Figure \ref{fig:model}a shows a schematic diagram of the flow field surrounding the droplet. 
Note that previous studies utilise an axial airflow to ventilate the accumulated vapour in the outer streaming vortices \citep{Yarin1999, Schiffter2007,Al2011}, allowing for direct use of the equations for inner convection (equation (\ref{average})). 
However, our study will focus on the effect of the entire flow field generated by the acoustic levitation method, including both inner and outer acoustic streaming, on the droplet evaporation. 

For the inner acoustic streaming region, the mass transfer of the acoustic streaming flow can be characterised by the average Sherwood number by \citet{Yarin1999}:

\begin{equation}
S\!\!\:h_{0,i}=1.89 \frac{B_i}{\left(\omega \cdot \mathcal{D}_{i}^g\right)^{1 / 2}}
\label{average}
\end{equation}

Here, $\omega$ is the angular frequency of the sound vibration. $\mathcal{D}_{i}^g$ the mass diffusion coefficient of the component $i$ in gas phase. $B_i=\frac{A_{0e,i}}{\rho_0 c_0}$ is an acoustic velocity scale of  component $i$, where $A_{0e,i}, \rho_0, c_0$ are the effective pressure amplitude of the acoustic field, unperturbed gas density and the sound velocity. The effective pressure amplitude $A_{0e,i}$ will significantly affect the evaporation rate of the droplet according to \citet{Junk2020}, which can also be derived from equation (\ref{average}). We have maintained stable value of $A_{0e}$ in our experiments, which can be obtained via equation (5.6) in \citet{Yarin1998} through obtaining the critical levitator power for dropout.  $A_{0e,w}=\unit{5521}{\newton/\meter^2}$ for water and $A_{0e,e}=\unit{5850}{\newton/\meter^2}$ for ethanol.

The heat transfer of the streaming flow is characterized by the Nusselt number calculated in the similar form:
\begin{equation}
N\!\!\:u_{0}=1.89 \frac{\overline{B}}{\left(\omega \cdot \alpha^g\right)^{1 / 2}}
\label{Nu}
\end{equation}
Here, ${\alpha}^g$ is the thermal diffusivity of air and $\overline{B}$ is the  acoustic velocity scale averaged according to component mass fraction in the liquid.

Due to the intense, rapid mixing in the vortices, the outer streaming region is assumed to have a uniform temperature $T_{outer}$ and uniform vapour concentration field $c_{i,outer}$ for species $i$. 
This can be justified by comparing the circulation time scale with the diffusion time according to \citet{Kronig1951}, i.e. by examining the heat and mass Peclet numbers. 
\begin{equation}
Pe_{mass}=\frac{Lu}{D}, Pe_{heat}=\frac{Lu}{\alpha},
\end{equation}
where the circulation velocity in the vortex $u$ is approximately $\unit{50}{\milli\meter/\second}$ according to \citet{Kobayashi2018}, the length scale of the vortex $L\approx \unit{10^{-2}}{\meter}$, the diffusion coefficient ${D}$ and the thermal diffusivity of the gas $\alpha$ are both of the order of $\unit{10^{-5}}{\meter^2/\second}$. Therefore, $Pe_{mass}, Pe_{heat}\approx50$ indicating that the advection is much faster than the diffusion, supporting our assumption.

Figure \ref{fig:model}(b) shows the distribution of the vapour mass concentration of component $i$ as a function of radius $r$.
The vapour concentration decreases rapidly from its peak at the gas-liquid interface, as a result of convection in the inner streaming region, and eventually reaches the outer streaming region where a uniform concentration field is established by the vortices. 
At the edge of the outer region, the vapour components diffuse into the far field. 
Here we assume that the outer streaming region is in an equilibrium stage for mass and heat transfer, where the mass and heat flow from the droplet to the outer streaming region is balanced by the mass and heat flow from the outer streaming region to the surroundings. 
From that, we derive the following equations for the mass flow $m_i$  (\ref{mass}) and heat flow $Q$ (\ref{heat}):
\begin{equation}
m_i=2 \pi R_d  D_0 {S\!\!\:h}_{0,i} \left(c_{s,i}-c_{\text{outer},i}\right)=4 \pi R^* D_0\left(c_{\text{outer},i}-c_{\infty,i}\right).
\label{mass}
\end{equation}
\begin{equation}
Q=2 \pi R_d  \lambda {N\!\!\:u}_{0} \left(T_s-T_{\text{outer}}\right)=4 \pi R^* \lambda\left(T_{\text{outer}}-T_{\infty}\right).
\label{heat}
\end{equation}

Defining $ k^*=R^*/R_d $, one can get the equations for $c_{\text{outer},i}$ and $T_{\text{outer}}$:
\begin{equation}
c_{\text{outer},i}=\frac{1}{1+\frac{2k^*}{S\!\!\:h_{0,i}}} c_{s,i}+\frac{1}{1+\frac{S\!\!\:h_{0,i}}{2k^*}} c_{\infty,i}, \ 
T_{\text{outer}}=\frac{1}{1+\frac{2k^*}{N\!\!\: u}} T_s+\frac{1}{1+\frac{N\!\!\: u}{2k^*}} T_{\infty},
\label{Tout}
\end{equation}
Substituting \ref{Tout} into \ref{mass} and \ref{heat}, one can eliminate the parameters of the outer region and obtain:
\begin{equation}
m_i=2 \pi R_d D_0 \frac{{S\!\!\: h}_{0,i}}{1+{S\!\!\: h}_{0,i} /2k^*}  \left(c_{s,i}-c_{\text{outer},i}\right), \ 
Q=2 \pi R_d \lambda  \frac{{N\!\!\: u}_{0}}{1+{N\!\!\: u}_{0} /2k^*}  \left(T_{s}-T_{\text{outer}}\right)
\end{equation}

From this, we can define the effective Sherwood and Nusselt numbers that consider the effect of outer acoustic streaming: 
\begin{equation}
{S\!\!\: h}_{e\!f\!\!f, i}=\frac{{S\!\!\: h}_{0,i}}{1+{S\!\!\: h}_{0,i} /2k^*}, {N\!\!\: u}_{e\!f\!\!f}=\frac{{N\!\!\: u}_{0}}{1+{N\!\!\: u}_{0} /2k^*}
\label{modified}
\end{equation}

We first obtain the Sherwood and Nusselt numbers induced by the inner acoustic streaming, ${S\!\!\: h}_{0,i}$ and  ${N\!\!\: u}_{0}$,  through (\ref{average}) and calculate the effective parameters accounting for the effects of the outer streaming, ${S\!\!\: h}_{e\!f\!\!f, i}$ and ${N\!\!\: u}_{e\!f\!\!f}$, through (\ref{modified}). Then, we implement the obtained ${S\!\!\: h}_{e\!f\!\!f, i}$ and ${N\!\!\: u}_{e\!f\!\!f}$ into the equations (\ref{interface}) and (\ref{temp}) for the numerical simulation. 

\section{Results and discussion}
\label{sec:results}
\subsection{Mass transfer}
\label{sec:mass}
\begin{figure}
\centering
\includegraphics[width=0.93\linewidth]{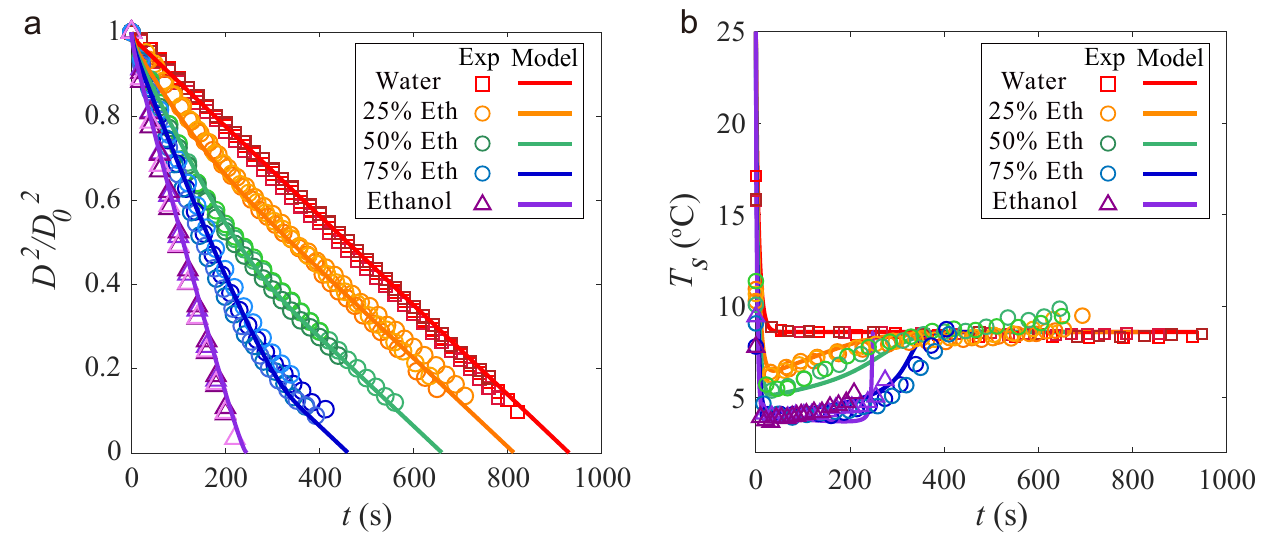}
  \caption{Normalised surface area $D^2/D_0^2$ (\textit{a}) and surface temperature $T_{s}$ (\textit{b})  versus time $t$ for various initial concentrations of the droplets with initial diameter square $D_0=\unit{1.414}{\milli\meter}$, evaporating under gas humidity $\rm{RH}\approx5\%$. Experimental results (markers) and model results (lines) are compared.}
\label{fig:expdata}
\end{figure}

To study the evaporation characteristics, we focus on the variation of the droplet volume by analysing the images captured by the CCD camera (see figure.~\ref{fig:setup}b). Figure \ref{fig:expdata}(\textit{a}) shows the temporal variation of the normalised surface area $D^2/D_0^2$ of droplets with different initial concentrations, where $D$ is the volume equivalent diameter of the droplet and $D_0$ is the initial value of $D$. The good agreement between the experimental (markers) and calculated results (solid curves) confirms the reliability of our model.

It is demonstrated that for pure water droplets, the diameter square $D^2$ remains linear with time under the influence of the acoustic field. This trend is consistent with the $d^2 -$ law, which focuses on the evaporation of pure, spherical droplets and states that the diameter square decreases linearly with time, at a rate determined by the ambient properties \citep{Finneran2021deviations, Sazhin2014droplets}. 
Since the slope of the $D^2/D_0^2$ curve  is directly proportional to the Sherwood number, the linear variation of the curve confirms the assumption that the original evaporation equation can be modified in the acoustic field using a constant Sherwood number via equation (\ref{MassDiff}). The Sherwood number of the inner acoustic streaming is approximately 15, calculated by equation (\ref{average}), leading to an evaporation in the inner acoustic streaming of $Sh/2$, i.e. 7.5 times faster than in a pure diffusion scenario. Nevertheless, the droplet lifetime is only reduced to half that of pure diffusive evaporation under the same ambient condition ($T_0=\unit{25}{\degree C}, RH=5\%$). This is due to the accumulation of vapour substances in the outer vortices, which reduces the effective Sherwood number ${S\!\!\: h}_{e\!f\!\!f, i}$ to about 3.
Therefore, both the inner and outer acoustic streaming have a significant effect on the evaporation. The equation (\ref{average}) cannot be used to calculate the evaporation alone if there is no ventilation flow to eliminate the outer acoustic streaming \citep{Yarin1999, Al2011}.

\begin{figure}
\centering
\includegraphics[width=1\linewidth]{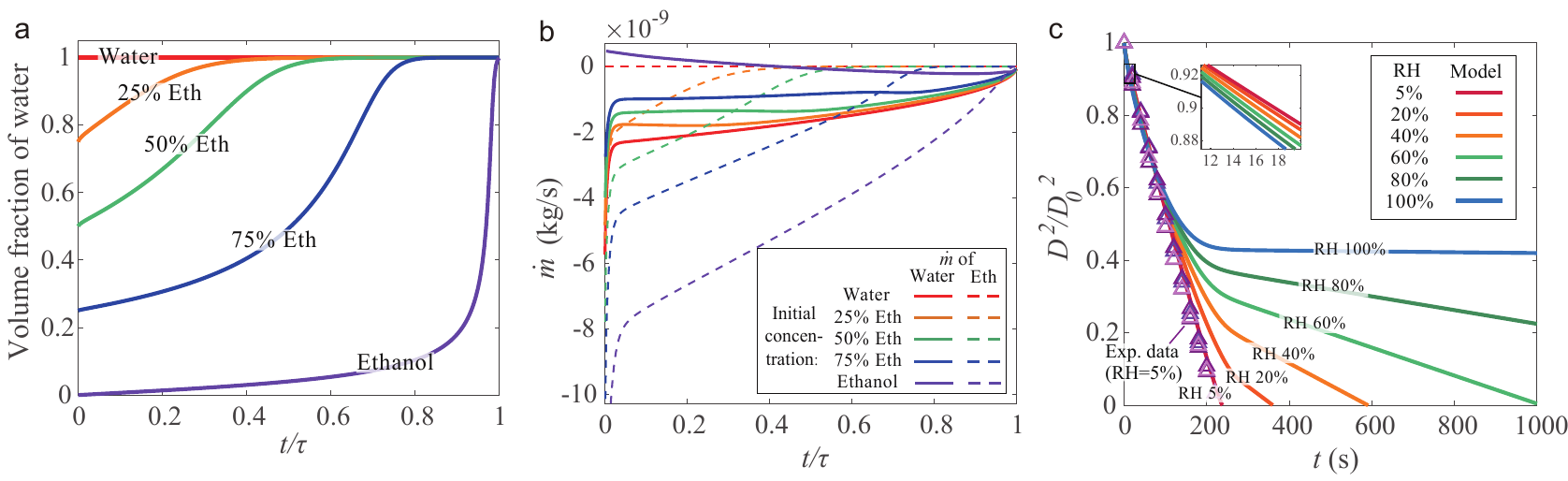}
  \caption{ (\textit{a}) Volume fraction of the water component in the droplet versus dimensionless time $t/\tau$, where $\tau$ is the calculated lifetime of the droplet. (\textit{b}) Temporal variation of the mass flow rate of water and ethanol with time for droplets with various initial ratios. (\textit{c}) Effect of ambient gas humidity on the variation of normalised surface area $D^2/D_0^2$. Model results of gas humidity varying from 5\% to 100\% and experimental results of RH=5\% are shown.}
\label{fig:eva-mod}
\end{figure}

The $D^2/D_0^2$ curves of pure ethanol droplets follow a similar trend to that of water droplets, with their slope exceeding that of water droplets due to the higher saturation pressure and molecular weight of ethanol. It is noticeable that, despite the low humidity, there is a slight bend at the end of the ethanol curve due to water condensation, which will be discussed in the subsequent paragraph.
In the case of bicomponent droplets, the evaporation curves display a noticeable shift from a steeper to a gentler slope, corresponding to the variation of the component fractions in the droplet.
From figure \ref{fig:eva-mod}(\textit{a}), we can see that the volume fraction of water increases during evaporation from the initial concentration to 1, regardless of the initial value. 
It can be obtained that the $D^2/D_0^2$ curves in figure \ref{fig:expdata}(\textit{a}) exhibit a transition point at the stage where water constitutes nearly 95\% of the total volume of the droplet.
Consequently, all droplets with varying initial concentrations become water droplets in the final stage of evaporation, thus the evaporation curves of the second stage run parallel to that of water droplets.

Focusing on the mass change rate of the two components in the evaporating droplets (see figure \ref{fig:eva-mod}(\textit{b})), it is evident that as the initial concentration of ethanol increases, the mass rate of the ethanol component increases while the mass rate of the water component decreases.
As the rise in the evaporation rate of water cannot compensate for the decline in the evaporation rate of ethanol, droplets evaporate faster at higher ethanol concentrations, as demonstrated in figure \ref{fig:expdata}(\textit{a}). 
It can also be seen that for bicomponent droplets, the evaporation rate of water is comparable to that of ethanol throughout the evaporation process, justifying that the role of water evaporation cannot be ignored, even if the water fraction is small. 
At extremely low initial water concentrations, particularly in the case of pure ethanol droplets, the interfacial vapour pressure of the water is insufficient to overcome the ambient vapour pressure, leading to condensation of the moisture on the droplet surface. Consequently, the mass rate of water in the ethanol droplet is initially positive, and evaporation losses are slightly offset.
If humidity level increases, condensation will intensify, resulting in a more complex trend of the $D^2$ curve \citep{Sasaki2020, Yang2023evaporation}. Figure \ref{fig:eva-mod}(\textit{c}) illustrates the effect of the ambient gas humidity on the evaporation rate. Model results demonstrate that there are distinct transitions in the slope of the $D^2$ curves, with curves having a similar trend in the first stage ($D^2/D_0^2>0.6$), and the slopes decreasing with increasing humidity in the second stage ($D^2/D_0^2<0.6$).
In addition, we find that the higher the ambient humidity, the faster the evaporation rate in the first stage ($D^2/D_0^2>0.6$), as depicted in the inset of figure \ref{fig:eva-mod}(\textit{c}). This can be attributed to the condensation of water compensating for the heat consumption caused by evaporative cooling, which enables the ethanol component to evaporate at a higher temperatures \citep{Li2023high}.

\subsection{Heat transfer}
Referring to heat transfer characteristics during the evaporation process, figure \ref{fig:expdata}(\textit{b}) displays the droplet surface temperature $T_s$ over time. It can be seen that a pure water droplet maintains a constant surface temperature ($\unit{\approx 8.5}{\degree C}$) after initially cooling down, signifying that it reaches an equilibrium between cooling by evaporation and heat transfer from the surrounding gas.
The temperature variation of the ethanol droplet follows a similar trend, with the surface temperature after initial cooling ($\unit{\approx 3.5}{\degree C}$) being lower than that of a water droplet due to its greater evaporation rate.  Furthermore, the ethanol droplet reaches the same surface temperature as a pure water droplet in the end, as a result of the condensation of water vapour onto the droplet. 
For bicomponent droplets, the surface temperature initially decreases rapidly to a minimum level, which is lower than that of a pure water droplet, before increasing slowly to eventually reach the same surface temperature as the water droplet.  
Increasing the ethanol concentration reduces the minimum temperature of the droplet, yet it is still higher than that of a pure ethanol droplet. 
Furthermore, the fact that all kinds of droplets reach a similar temperature at the end is consistent with our conclusion in section \ref{sec:mass} that water becomes the dominant component at the end of the evaporation process, regardless of the initial concentration ratio.

\begin{figure}
\centering
\includegraphics[width=0.5\linewidth]{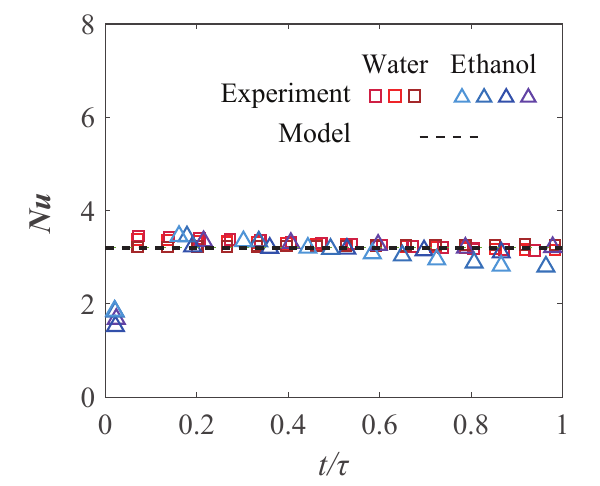}
  \caption{Nusselt number $Nu$ calculated through experimental results (markers) and model (dashed line) for pure water and ethanol droplets. $\tau$ relates to the lifetime of the droplet.}
\label{fig:Nu}
\end{figure}

From our experimental results, we are able to calculate the heat transfer coefficient for an evaporating pure droplet using the energy conservation law:
\begin{equation}
\rho_L c_p \frac{d\left(V T_{\text {s}}\right)}{d t}=h A\left(T_{\infty}-T_{\text {s}}\right)+\rho_L L \frac{d V}{d t}
\label{energy}
\end{equation}

Here, $\rho_L$ is the density of the liquid, $c_p$ is the specific heat capacity at constant temperature, $L$ the latent heat of the liquid and $T_{\infty}$ the surrounding gas temperature.
$V, A, T_s$ are the droplet volume, droplet surface area, and surface temperature, respectively, and can be obtained from experimental results. Here we assume a uniform temperature field within the droplet, which is supported by our numerical results discussed in Section \ref{distribution}.
Equation (\ref{energy}) can be solved for the heat transfer coefficient $h$, and thus the Nusselt number $Nu$ can be calculated by the following equation:
\begin{equation}
Nu=\frac{2 R_d h}{\lambda}=\frac{2 R_d \rho_L \left(c_p \frac{d\left(V T_{\text {s}}\right)}{d t}-L \frac{d V}{d t}\right)}{\lambda A\left(T_{\infty}-T_{\text {s}}\right)}
\label{h}
\end{equation}
Here, $R_d=D/2$ is the volume equivalent radius of the droplet.
When calculating the Nusselt number for the evaporation of ethanol droplets, we neglect the condensation of water vapour because at such a low humidity (RH\textcolor{blue}{=}5\%), the role of condensation heat is very small.
Figure \ref{fig:eva-mod}(\textit{c}) shows the temporal variation of $Nu$ for water and ethanol droplets calculated directly from experimental data (markers) with the model calculation (dashed line). 
Except for the initial stage, the Nusselt number remains approximately constant during evaporation. The experimental results calculated via equation (\ref{modified}) are in agreement with the experimental data calculated via equation (\ref{h}), demonstrating the validity of our model.

\begin{figure}
\centering
\includegraphics[width=0.8\linewidth]{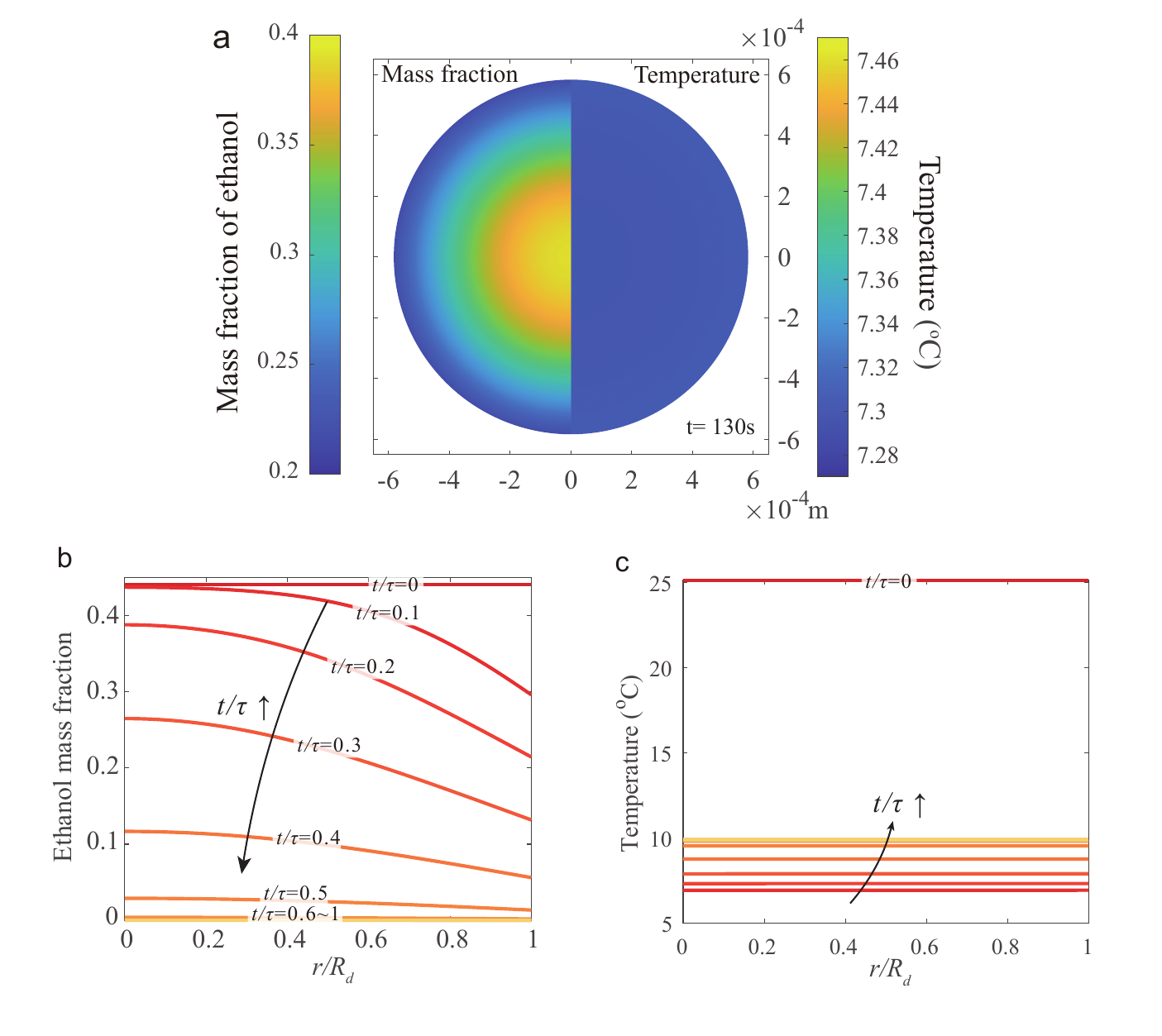}
  \caption{\textbf{For droplet with initial ethanol volume fraction of 50 \%:} (\textit{a}) Distribution of ethanol mass fraction and temperature of the droplet at $t=\unit{130}{\second}$ ($t/\tau=0.2$). Temporal variation of ethanol mass fraction (\textit{b})  and temperature distribution (\textit{c}) within the droplet as a function of dimensionless radius $r/R_d$. $R_d$ is the equavalent radius of the droplet.}
\label{fig:moddata}
\end{figure}

\subsection{Temperature and concentration distribution}
\label{distribution}
With the calculated results, we are able to obtain specific parameter distributions within the droplet.
Figure \ref{fig:moddata}(\textit{a}) displays the distribution of ethanol concentration and liquid temperature for a droplet with an initial ethanol volume fraction of 50\% at $t=\unit{130}{\second}$. The complete variation of the concentration and temperature field during the evaporation process is displayed in the supplementary video.
The temporal variation of the concentration and temperature fields within the droplet is shown in figures \ref{fig:moddata}(\textit{b}) and (\textit{c}), respectively.
It is evident that the temperature maintains uniformity throughout the evaporation (with a temperature difference of less than $\unit{0.01}{\degree C}$), while the concentration field shows significant inhomogeneity near the droplet interface at the start, with the gradient decreasing over time. 
We use the Biot number to study the uniformity of the internal concentration and temperature fields during evaporation.
According to the definition, there exists a relationship between the mass transfer Biot number $Bi_m$ and the heat transfer Biot number $Bi_q$ with Sherwood and Nusselt numbers:
$
Bi_m=Sh\frac{D_g}{D_l}, Bi_q=Nu\frac{k_g}{k_l}.
$
$Bi_m$ is calculated to be of the order of $10^2$, indicating that evaporation losses at the droplet interface cannot be compensated by species transport from the interior of the droplet, resulting in an inhomogeneous concentration field. $Bi_q$, on the other hand, is of the order of 0.1, allowing the temperature field to remain uniform at all times. 
Comparing the characteristic time scales of heat transfer and component diffusion with the time scale of droplet evaporation also gives consistent conclusions \citep{Yarin2002}.

As a result, the non-uniformity of the concentration field within the droplet must be taken into account in the modelling, while the temperature field can be considered homogeneous to simplify the calculation. 
However, our model does not take into account the effect caused by the internal circulation flow inside the droplet, so the generality of this conclusion needs to be further demonstrated.

\section{Conclusions}
In this study, we carried out a systematic measurement of the evaporation characteristics of acoustically levitated droplets with varying initial concentrations. 
The evaporation process was modelled using an in-house finite element method framework and included the effects of both convective flow of the inner acoustic region and vortices of the outer acoustic region.
By comparing the experimental and numerical results, it was shown that constant Sherwood and Nusselts numbers for mass and heat transfer can describe droplet evaporation under acoustic levitation. 
Both the inner and outer acoustic streaming contribute considerable to the mass and heat transfer, resulting in the effective parameters ${S\!\!\: h}_{e\!f\!\!f, i}, {N\!\!\: u}_{e\!f\!\!f}\approx3$.
Additionally, the analysis of the concentration and temperature fields within the droplet confirmed the importance of considering the concentration gradient within the droplet.

In forthcoming research, it is intended that the acoustic field equations should replace the analytical solutions in order to fully calculate the flow field around the droplet. The numerical simulation should be extended to two or three dimensions for a thorough assessment of the internal flow of the droplet.
Additionally, explorations could focus on immiscible components and phase separation processes to enhance practical applications \citep{Lyu2021explosive, tan2016evaporation, Diddens2017evaporating}.

\vspace{-4 mm}

\section*{Acknowledgements}
We thank D. Lohse, H. Tan and S. Lyu for insightful discussions. This work is supported by National Natural Science Foundation of China under Grants Nos. 11988102, 51976105, 91841302, the National Key R\&D Program of China (Grant No. 2021YFA0716201) and the New Cornerstone Science Foundation through the New Cornerstone Investigator Program and the XPLORER PRIZE.

\vspace{-4 mm}
\section*{Declaration of interests}
The authors report no conflict of interest.



\begin{thebibliography}{50}
\expandafter\ifx\csname natexlab\endcsname\relax\def\natexlab#1{#1}\fi
\def\au#1{#1} \def\ed#1{#1} \def\yr#1{#1}\def\at#1{#1}\def\jt#1{\textit{#1}}
  \def\bt#1{#1}\def\bvol#1{\textbf{#1}} \def\vol#1{#1} \def\pg#1{#1}
  \def\publ#1{#1}\def\arxiv#1{#1}\def\org#1{#1}\def\st#1{\textit{#1}}

\bibitem[Al~Zaitone(2018)]{Al2018}
{\sc \au{Al~Zaitone, B.}} \yr{2018}  \at{Oblate spheroidal droplet evaporation
  in an acoustic levitator}.  \jt{Int. J. Heat Mass Transfer}  \bvol{126},
  \pg{164--172}.

\bibitem[Al~Zaitone \& Tropea(2011)]{Al2011}
{\sc \au{Al~Zaitone, B.} \& \au{Tropea, C.}} \yr{2011}  \at{Evaporation of pure
  liquid droplets: Comparison of droplet evaporation in an acoustic field
  versus glass-filament}.  \jt{Chem. Eng. Sci.}  \bvol{66}~(17),
  \pg{3914--3921}.

\bibitem[Andrade {\em et~al.\/}(2018)Andrade, P{\'e}rez \&
  Adamowski]{Andrade2018Review}
{\sc \au{Andrade, M.A.B.}, \au{P{\'e}rez, N.} \& \au{Adamowski, J.C.}}
  \yr{2018}  \at{Review of progress in acoustic levitation}.  \jt{Braz. J.
  Phys.}  \bvol{48},  \pg{190--213}.

\bibitem[B{\"a}nsch \& G{\"o}tz(2018)]{Bansch2018numerical}
{\sc \au{B{\"a}nsch, E.} \& \au{G{\"o}tz, M.}} \yr{2018}  \at{Numerical study
  of droplet evaporation in an acoustic levitator}.  \jt{Phys. Fluids}
  \bvol{30}~(3),  \pg{037103}.

\bibitem[Brenn {\em et~al.\/}(2007)Brenn, Deviprasath, Durst \&
  Fink]{Brenn2007}
{\sc \au{Brenn, G.}, \au{Deviprasath, L.J.}, \au{Durst, F.} \& \au{Fink, C.}}
  \yr{2007}  \at{Evaporation of acoustically levitated multi-component liquid
  droplets}.  \jt{Int. J. Heat Mass Transfer}  \bvol{50}~(25-26),
  \pg{5073--5086}.

\bibitem[Chen {\em et~al.\/}(2022)Chen, Li, Zhang, Zhang \& Zang]{Chen2022}
{\sc \au{Chen, H.}, \au{Li, A.}, \au{Zhang, Y.}, \au{Zhang, X.} \& \au{Zang,
  D.}} \yr{2022}  \at{Evaporation and liquid-phase separation of
  ethanol--cyclohexane binary drops under acoustic levitation}.  \jt{Phys.
  Fluids}  \bvol{34}~(9),  \pg{092108}.

\bibitem[Diddens(2017)]{Diddens2017detailed}
{\sc \au{Diddens, C.}} \yr{2017}  \at{Detailed finite element method modeling
  of evaporating multi-component droplets}.  \jt{J. Comput. Phys.}  \bvol{340},
   \pg{670--687}.

\bibitem[Diddens {\em et~al.\/}(2021)Diddens, Li \&
  Lohse]{Diddens2021competing}
{\sc \au{Diddens, C.}, \au{Li, Y.} \& \au{Lohse, D.}} \yr{2021}  \at{Competing
  marangoni and rayleigh convection in evaporating binary droplets}.  \jt{J.
  Fluid Mech.}  \bvol{914},  \pg{A23}.

\bibitem[Diddens {\em et~al.\/}(2017)Diddens, Tan, Lv, Versluis, Kuerten, Zhang
  \& Lohse]{Diddens2017evaporating}
{\sc \au{Diddens, C.}, \au{Tan, H.}, \au{Lv, P.}, \au{Versluis, M.},
  \au{Kuerten, J.G.M.}, \au{Zhang, X.} \& \au{Lohse, D.}} \yr{2017}
  \at{Evaporating pure, binary and ternary droplets: thermal effects and axial
  symmetry breaking}.  \jt{J. Fluid Mech.}  \bvol{823},  \pg{470--497}.

\bibitem[Doss \& B{\"a}nsch(2022)]{Doss2022}
{\sc \au{Doss, M.} \& \au{B{\"a}nsch, E.}} \yr{2022}  \at{Numerical study of
  single droplet drying in an acoustic levitator before the critical point of
  time}.  \jt{Chem. Eng. Sci.}  \bvol{248},  \pg{117149}.

\bibitem[Finneran {\em et~al.\/}(2021)Finneran, Garner \&
  Nadal]{Finneran2021deviations}
{\sc \au{Finneran, J.}, \au{Garner, C.P.} \& \au{Nadal, F.}} \yr{2021}
  \at{Deviations from classical droplet evaporation theory}.  \jt{Proc. R. Soc.
  A}  \bvol{477}~(2251),  \pg{20210078}.

\bibitem[Fuchs(1959)]{Fuchs1959}
{\sc \au{Fuchs, N.A.}} \yr{1959} {\em Evaporation and Droplet Growth in Gaseous
  Media\/}.  \publ{Pergamon Press, London}.

\bibitem[Heil \& Hazel(2006)]{Heil2006}
{\sc \au{Heil, M.} \& \au{Hazel, A.L.}} \yr{2006}  \at{{oomph-lib - An
  Object-oriented multi-physics finite-element library}}.  \jt{Lect. Notes
  Comput. Sci. Eng.}  \bvol{53},  \pg{19--49}.

\bibitem[Junk {\em et~al.\/}(2020)Junk, Hinrichs, Polt, Fechner \&
  Pauer]{Junk2020}
{\sc \au{Junk, M.}, \au{Hinrichs, J.}, \au{Polt, F.}, \au{Fechner, J.} \&
  \au{Pauer, W.}} \yr{2020}  \at{Quantitative experimental determination of
  evaporation influencing factors in single droplet levitation}.  \jt{Int. J.
  Heat Mass Transfer}  \bvol{149},  \pg{119057}.

\bibitem[Kobayashi {\em et~al.\/}(2018)Kobayashi, Goda, Hasegawa \&
  Abe]{Kobayashi2018}
{\sc \au{Kobayashi, K.}, \au{Goda, A.}, \au{Hasegawa, K.} \& \au{Abe, Y.}}
  \yr{2018}  \at{Flow structure and evaporation behavior of an acoustically
  levitated droplet}.  \jt{Phys. Fluids}  \bvol{30}~(8),  \pg{082105}.

\bibitem[Kronig \& Brink(1951)]{Kronig1951}
{\sc \au{Kronig, R.} \& \au{Brink, J.C.}} \yr{1951}  \at{On the theory of
  extraction from falling droplets}.  \jt{Appl. Sci. Res.}  \bvol{2},
  \pg{142--154}.

\bibitem[Law \& Law(1982)]{Law1982}
{\sc \au{Law, C.K.} \& \au{Law, H.K.}} \yr{1982}  \at{A d2-law for
  multicomponent droplet vaporization and combustion}.  \jt{AIAA J.}
  \bvol{20}~(4),  \pg{522--527}.

\bibitem[Lee \& Wang(1990)]{Lee1990}
{\sc \au{Lee, C.P.} \& \au{Wang, T.G.}} \yr{1990}  \at{Outer acoustic
  streaming}.  \jt{J. Acoust. Soc. Am.}  \bvol{88}~(5),  \pg{2367--2375}.

\bibitem[Li {\em et~al.\/}(2023)Li, Lohse \& Huisman]{Li2023high}
{\sc \au{Li, M.}, \au{Lohse, D.} \& \au{Huisman, S.G.}} \yr{2023}  \at{High
  humidity enhances the evaporation of non-aqueous volatile sprays}.  \jt{J.
  Fluid Mech.}  \bvol{956},  \pg{A19}.

\bibitem[Lieber {\em et~al.\/}(2021)Lieber, Melekidis, Koch \&
  Bauer]{Lieber2021}
{\sc \au{Lieber, C.}, \au{Melekidis, S.}, \au{Koch, R.} \& \au{Bauer, H.}}
  \yr{2021}  \at{Insights into the evaporation characteristics of saliva
  droplets and aerosols: Levitation experiments and numerical modeling}.
  \jt{J. Aerosol Sci.}  \bvol{154},  \pg{105760}.

\bibitem[Lierke(1996)]{Lierke1996}
{\sc \au{Lierke, E.G.}} \yr{1996}  \at{Akustische positionierung-ein
  unfassender uberblick {\"u}ber grundlagen und anwendungen}.  \jt{Acustica}
  \bvol{82}~(2),  \pg{220--237}.

\bibitem[Lohse(2022)]{Lohse2022fundamental}
{\sc \au{Lohse, D.}} \yr{2022}  \at{Fundamental fluid dynamics challenges in
  inkjet printing}.  \jt{Annu. Rev. Fluid Mech.}  \bvol{54},  \pg{349--382}.

\bibitem[Lyu {\em et~al.\/}(2021)Lyu, Tan, Wakata, Yang, Law, Lohse \&
  Sun]{Lyu2021explosive}
{\sc \au{Lyu, S.}, \au{Tan, H.}, \au{Wakata, Y.}, \au{Yang, X.}, \au{Law,
  C.K.}, \au{Lohse, D.} \& \au{Sun, C.}} \yr{2021}  \at{On explosive boiling of
  a multicomponent leidenfrost drop}.  \jt{Proc. Natl. Acad. Sci. U. S. A.}
  \bvol{118}~(2),  \pg{e2016107118}.

\bibitem[Maqua {\em et~al.\/}(2008)Maqua, Castanet, Grisch, Lemoine, Kristyadi
  \& Sazhin]{Maqua2008Monodisperse}
{\sc \au{Maqua, C.}, \au{Castanet, G.}, \au{Grisch, F.}, \au{Lemoine, F.},
  \au{Kristyadi, T.} \& \au{Sazhin, S.S.}} \yr{2008}  \at{Monodisperse droplet
  heating and evaporation: experimental study and modelling}.  \jt{Int. J. Heat
  Mass Transfer}  \bvol{51}~(15-16),  \pg{3932--3945}.

\bibitem[Maxwell(1877)]{Maxwell1877}
{\sc \au{Maxwell, J.C.}} \yr{1877}  \at{Diffusion}.  \jt{Encyclopaedia
  britannica, 9th edn}  \bvol{7},  \pg{214–221}.

\bibitem[O'Connell {\em et~al.\/}(2023)O'Connell, Sharratt \&
  Cabral]{OConnell2023}
{\sc \au{O'Connell, R.A.}, \au{Sharratt, W.N.} \& \au{Cabral, J.T.}} \yr{2023}
  \at{Breath figure assembly on evaporating polymer solution droplets in
  levitation}.  \jt{Phys. Rev. Lett.}  \bvol{131},  \pg{218101}.

\bibitem[Riley(2001)]{Riley2001}
{\sc \au{Riley, N.}} \yr{2001}  \at{Steady streaming}.  \jt{Annu. Rev. Fluid
  Mech.}  \bvol{33}~(1),  \pg{43--65}.

\bibitem[Sasaki {\em et~al.\/}(2020)Sasaki, Hasegawa, Kaneko \&
  Abe]{Sasaki2020}
{\sc \au{Sasaki, Y.}, \au{Hasegawa, K.}, \au{Kaneko, A.} \& \au{Abe, Y.}}
  \yr{2020}  \at{Heat and mass transfer characteristics of binary droplets in
  acoustic levitation}.  \jt{Phys. Fluids}  \bvol{32}~(7),  \pg{072102}.

\bibitem[Sasaki {\em et~al.\/}(2019)Sasaki, Kobayashi, Hasegawa, Kaneko \&
  Abe]{Sasaki2019}
{\sc \au{Sasaki, Y.}, \au{Kobayashi, K.}, \au{Hasegawa, K.}, \au{Kaneko, A.} \&
  \au{Abe, Y.}} \yr{2019}  \at{Transition of flow field of acoustically
  levitated droplets with evaporation}.  \jt{Phys. Fluids}  \bvol{31}~(10),
  \pg{102109}.

\bibitem[Sazhin(2014)]{Sazhin2014droplets}
{\sc \au{Sazhin, S.}} \yr{2014} {\em Droplets and sprays\/}.  \publ{Springer}.

\bibitem[Schiffter \& Lee(2007)]{Schiffter2007}
{\sc \au{Schiffter, H.} \& \au{Lee, G.}} \yr{2007}  \at{Single-droplet
  evaporation kinetics and particle formation in an acoustic levitator. part 1:
  Evaporation of water microdroplets assessed using boundary-layer and acoustic
  levitation theories}.  \jt{J. Pharm. Sci.}  \bvol{96}~(9),  \pg{2274--2283}.

\bibitem[Shi \& Apfel(1996)]{Shi1996}
{\sc \au{Shi, W.T.} \& \au{Apfel, R.E.}} \yr{1996}  \at{Deformation and
  position of acoustically levitated liquid drops}.  \jt{J. Acoust. Soc. Am.}
  \bvol{99}~(4),  \pg{1977--1984}.

\bibitem[Sirignano(1983)]{Sirignano1983}
{\sc \au{Sirignano, W.A.}} \yr{1983}  \at{Fuel droplet vaporization and spray
  combustion theory}.  \jt{Prog. Energy Combust. Sci.}  \bvol{9}~(4),
  \pg{291--322}.

\bibitem[Sirignano(2010)]{Sirignano2010fluid}
{\sc \au{Sirignano, W.A.}} \yr{2010} {\em Fluid dynamics and transport of
  droplets and sprays\/}.  \publ{Cambridge University Press}.

\bibitem[Tan {\em et~al.\/}(2016)Tan, Diddens, Lv, Kuerten, Zhang \&
  Lohse]{tan2016evaporation}
{\sc \au{Tan, H.}, \au{Diddens, C.}, \au{Lv, P.}, \au{Kuerten, J.G.M.},
  \au{Zhang, X.} \& \au{Lohse, D.}} \yr{2016}  \at{Evaporation-triggered
  microdroplet nucleation and the four life phases of an evaporating ouzo
  drop}.  \jt{Proc. Natl. Acad. Sci. U. S. A.}  \bvol{113}~(31),
  \pg{8642--8647}.

\bibitem[Tian {\em et~al.\/}(1993)Tian, Holt \& Apfel]{Tian1993}
{\sc \au{Tian, Y.}, \au{Holt, R.G.} \& \au{Apfel, R.E.}} \yr{1993}
  \at{Deformation and location of an acoustically levitated liquid drop}.
  \jt{J. Acoust. Soc. Am.}  \bvol{93}~(6),  \pg{3096--3104}.

\bibitem[Tonini \& Cossali(2012)]{Tonini2012analytical}
{\sc \au{Tonini, S.} \& \au{Cossali, G.E.}} \yr{2012}  \at{An analytical model
  of liquid drop evaporation in gaseous environment}.  \jt{Int. J. Therm. Sci.}
   \bvol{57},  \pg{45--53}.

\bibitem[Tonini \& Cossali(2013)]{Tonini2013exact}
{\sc \au{Tonini, S.} \& \au{Cossali, G.E.}} \yr{2013}  \at{An exact solution of
  the mass transport equations for spheroidal evaporating drops}.  \jt{Int. J.
  Heat Mass Transfer}  \bvol{60},  \pg{236--240}.

\bibitem[Tonini \& Cossali(2019)]{Tonini2019analytical}
{\sc \au{Tonini, S.} \& \au{Cossali, G.E.}} \yr{2019}  \at{An analytical
  approach to model heating and evaporation of multicomponent ellipsoidal
  drops}.  \jt{Heat Mass Transfer}  \bvol{55},  \pg{1257--1269}.

\bibitem[Trinh \& Robey(1994)]{Trinh1994}
{\sc \au{Trinh, E.H.} \& \au{Robey, J.L.}} \yr{1994}  \at{Experimental study of
  streaming flows associated with ultrasonic levitators}.  \jt{Phys. Fluids}
  \bvol{6}~(11),  \pg{3567--3579}.

\bibitem[Wang {\em et~al.\/}(2022)Wang, Orejon, Takata \& Sefiane]{Wang2022}
{\sc \au{Wang, Z.}, \au{Orejon, D.}, \au{Takata, Y.} \& \au{Sefiane, K.}}
  \yr{2022}  \at{Wetting and evaporation of multicomponent droplets}.
  \jt{Phys. Rep.}  \bvol{960},  \pg{1--37}.

\bibitem[Yamamoto {\em et~al.\/}(2008)Yamamoto, Abe, Fujiwara, Hasegawa \&
  Aoki]{Yamamoto2008}
{\sc \au{Yamamoto, Y.}, \au{Abe, Y.}, \au{Fujiwara, A.}, \au{Hasegawa, K.} \&
  \au{Aoki, K.}} \yr{2008}  \at{Internal flow of acoustically levitated
  droplet}.  \jt{Microgravity Sci. Technol.}  \bvol{20},  \pg{277--280}.

\bibitem[Yang {\em et~al.\/}(2023)Yang, Pahlavan, Stone \&
  Bain]{Yang2023evaporation}
{\sc \au{Yang, L.}, \au{Pahlavan, A.A.}, \au{Stone, H.A.} \& \au{Bain, C.D.}}
  \yr{2023}  \at{Evaporation of alcohol droplets on surfaces in moist air}.
  \jt{Proc. Natl. Acad. Sci. U. S. A.}  \bvol{120}~(38),  \pg{e2302653120}.

\bibitem[Yarin {\em et~al.\/}(1999)Yarin, Brenn, Kastner, Rensink \&
  Tropea]{Yarin1999}
{\sc \au{Yarin, A.L.}, \au{Brenn, G.}, \au{Kastner, O.}, \au{Rensink, D.} \&
  \au{Tropea, C.}} \yr{1999}  \at{Evaporation of acoustically levitated
  droplets}.  \jt{J. Fluid Mech.}  \bvol{399},  \pg{151--204}.

\bibitem[Yarin {\em et~al.\/}(2002)Yarin, Brenn \& Rensink]{Yarin2002}
{\sc \au{Yarin, A.L.}, \au{Brenn, G.} \& \au{Rensink, D.}} \yr{2002}
  \at{Evaporation of acoustically levitated droplets of binary liquid
  mixtures}.  \jt{Int. J. Heat Fluid Flow}  \bvol{23}~(4),  \pg{471--486}.

\bibitem[Yarin {\em et~al.\/}(1998)Yarin, Pfaffenlehner \& Tropea]{Yarin1998}
{\sc \au{Yarin, A.L.}, \au{Pfaffenlehner, M.} \& \au{Tropea, C.}} \yr{1998}
  \at{On the acoustic levitation of droplets}.  \jt{J. Fluid Mech.}
  \bvol{356},  \pg{65--91}.

\bibitem[Zang {\em et~al.\/}(2019)Zang, Tarafdar, Tarasevich, Choudhury \&
  Dutta]{Zang2019}
{\sc \au{Zang, D.}, \au{Tarafdar, S.}, \au{Tarasevich, Y.Y.}, \au{Choudhury,
  M.D.} \& \au{Dutta, T.}} \yr{2019}  \at{Evaporation of a droplet: From
  physics to applications}.  \jt{Phys. Rep.}  \bvol{804},  \pg{1--56}.

\bibitem[Zang {\em et~al.\/}(2017)Zang, Yu, Chen, Li, Wu \&
  Geng]{Zang2017Acoustic}
{\sc \au{Zang, D.}, \au{Yu, Y.}, \au{Chen, Z.}, \au{Li, X.}, \au{Wu, H.} \&
  \au{Geng, X.}} \yr{2017}  \at{Acoustic levitation of liquid drops: Dynamics,
  manipulation and phase transitions}.  \jt{Adv. Colloid Interface Sci.}
  \bvol{243},  \pg{77--85}.

\bibitem[Zeng {\em et~al.\/}(2023)Zeng, Wakata, Chao, Li \& Sun]{Zeng2023}
{\sc \au{Zeng, H.}, \au{Wakata, Y.}, \au{Chao, X.}, \au{Li, M.} \& \au{Sun,
  C.}} \yr{2023}  \at{On evaporation dynamics of an acoustically levitated
  multicomponent droplet: Evaporation-triggered phase transition and freezing}.
   \jt{J. Colloid Interface Sci.}  \bvol{648},  \pg{736--744}.

\bibitem[Zuend {\em et~al.\/}(2008)Zuend, Marcolli, Luo \& Peter]{Zuend2008}
{\sc \au{Zuend, A.}, \au{Marcolli, C.}, \au{Luo, B.P.} \& \au{Peter, T.}}
  \yr{2008}  \at{A thermodynamic model of mixed organic-inorganic aerosols to
  predict activity coefficients}.  \jt{Atmos. Chem. Phys.}  \bvol{8}~(16),
  \pg{4559--4593}.

\end{thebibliography}

\end{document}